\documentclass{jpp}
\usepackage{graphicx}
\usepackage{xcolor}
\usepackage[utf8]{inputenc}
\usepackage[T1]{fontenc}
\usepackage{amsmath}
\usepackage{csquotes}
\usepackage{upgreek}
\usepackage{lineno}

\shorttitle{Eulerian and Lagrangian electron energization}
\shortauthor{K. Steinvall, L. Richard, T. F\"ul\"op, L. Hanebring, and I. Pusztai}

\title{Eulerian and Lagrangian electron energization during magnetic reconnection}

\author{Konrad Steinvall\aff{1}
  \corresp{\email{konrad.steinvall@chalmers.se}},
  Louis Richard\aff{2},
  T\"unde F\"ul\"op\aff{1},
 Lise Hanebring\aff{1},
 \and Istv\'an Pusztai\aff{1}}

\affiliation{\aff{1}Department of Physics, Chalmers University of Technology, Gothenburg, Sweden
\aff{2}Swedish Institute of Space Physics, Uppsala, Sweden}

\begin{document}

\maketitle
\begin{abstract}
% Max 250 words
Electron energization by magnetic reconnection has historically been studied in the Lagrangian guiding-center framework. Insights from such studies include that Fermi acceleration in magnetic islands can accelerate electrons to high energies. An alternative Eulerian fluid formulation of electron energization was recently used to study electron energization during magnetic reconnection in the absence of magnetic islands.
Here, we use particle-in-cell simulations to compare the Eulerian and Lagrangian models of electron energization in a setup where reconnection leads to magnetic island formation. 
We find the largest energization at the edges of magnetic islands. There, energization related to the diamagnetic drift dominates in the Eulerian model, while the Fermi related term dominates in the Lagrangian model. The models predict significantly different energization rates locally. A better agreement is found after integrating over the simulation domain. We show that strong magnetic curvature can break the magnetic moment conservation assumed by the Lagrangian model, leading to erroneous results.
The Eulerian fluid model is a complete fluid description and accurately models bulk energization. However, local measurements of its constituent energization terms need not reflect locations where plasma is heated or accelerated. The Lagrangian guiding center model can accurately describe the energization of particles, but it cannot describe the evolution of the fluid energy. We conclude that while both models can be valid, they describe two fundamentally different quantities, and care should be taken when choosing which model to use.
\end{abstract}

\section{Introduction}

Magnetic reconnection is responsible for explosive energy release in space and laboratory plasmas \citep{yamada2010,gonzalez2016}. By converting energy stored in the magnetic field to plasma energy, reconnection is able to accelerate electrons to high energies
\citep[e.g.][]{oieroset2002,drake2005,fu2013,oka2023}. The energization of electrons during magnetic reconnection has primarily been studied in the Lagrangian guiding center framework \citep{northorp1963}, where the electron motion around the guiding center is reduced to a magnetic moment $\mu=m_ev_\perp^2/(2B)$, which is assumed to be conserved along collisionless orbits \citep[e.g.][]{drake2006,dahlin2014,dahlin2015}. Here, $m_e$ is the electron mass, and $v_\perp$ is the velocity component perpendicular to the local magnetic field $\boldsymbol{B}$ that has the magnitude $B=|\boldsymbol{B}|$. Under this assumption, one can derive a differential equation for the evolution of a single guiding center's energy $\epsilon$, as
\begin{equation}
\label{eq:single_guiding}
    \frac{d\epsilon}{dt}=\mu\frac{\partial B}{\partial t} + q_e\left(v_\parallel\boldsymbol{b}+\boldsymbol{v}_g+\boldsymbol{v}_c\right)\cdot\boldsymbol{E},
\end{equation}
where $\boldsymbol{b}=\boldsymbol{B}/B$, $q_e$ is the electron charge, $v_\parallel$ is the velocity component parallel to the magnetic field, $\boldsymbol{v}_g$ and $\boldsymbol{v}_c$ are respectively the magnetic field gradient and curvature drifts, and $\boldsymbol{E}$ is the electric field \cite[e.g.][]{northorp1963, dahlin2014}.
The standard procedure is then to sum Eq.~(\ref{eq:single_guiding}) over an ensemble of guiding centers in a local region to derive a fluid-like equation describing the evolution of the total electron kinetic energy density $\mathcal{E}_\text{LGCM}$, 
\begin{equation}
\label{eq:guiding}
    \frac{d\mathcal{E}_\text{LGCM}}{dt}= J_{e\parallel}E_\parallel + \frac{p_{e\perp}}{B}\left(\frac{\partial B}{\partial t}+\boldsymbol{u}_{E}\cdot\nabla B\right)+\left(p_{e\parallel}+m_en_eu_{e\parallel}^2\right)\boldsymbol{u}_E\cdot\boldsymbol{k},
\end{equation}
where $p_{e\perp}$ and $p_{e\parallel}$ are respectively the electron pressure components perpendicular and parallel to the local $\boldsymbol{B}$, $J_{e\parallel}$ is the parallel component of the electron current density $\boldsymbol{J}_e=q_en_e\boldsymbol{u}_e$, where $n_e=\int d^3v f_e$ is the electron density, with the electron velocity distribution function (VDF) $f_e=f_e(\boldsymbol{r},\boldsymbol{v},t)$; $\boldsymbol{u}_e=n_e^{-1}\int d^3v \boldsymbol{v} f_e$ is the bulk electron velocity; $u_{e\parallel}$ is the component of $\boldsymbol{u}$ parallel to $\boldsymbol{B}$; $\boldsymbol{u}_E=\boldsymbol{E}\times\boldsymbol{B}/B^2$ is the E-cross-B-drift; and $\boldsymbol{k}=\boldsymbol{b}\cdot\nabla\boldsymbol{b}$ is the magnetic curvature. We will use the notation from \citet{tenbarge2024} to refer to the first, second, and third terms on the right-hand-side as $W_\parallel$, $W_\text{beta-gradB}$, and $W_\text{curv0}$ respectively. The first term is related to acceleration by parallel electric fields, the second to betatron acceleration, and the third to Fermi acceleration due to the curvature drift. We will refer to this model as the Lagrangian guiding center model (LGCM). Eq.\ (\ref{eq:guiding}) has been used extensively in both spacecraft data analysis \citep[e.g.][]{eriksson2020,zhong2020,jiang2021} and numerical studies \citep[e.g.][]{dahlin2014,dahlin2015,dahlin2016,li2015} to quantify electron energization during magnetic reconnection.

However, recent work by \citet{tenbarge2024} has shown that there are some often overlooked nuances in Eq.~(\ref{eq:guiding}). In particular, it is derived in a Lagrangian framework but often directly applied to an Eulerian grid. In their study, \citet{tenbarge2024} used a continuum Vlasov-Maxwell solver to simulate single x-point reconnection and investigate the Lagrangian model of electron energization. Their results showed that while the right-hand side of Eq.~(\ref{eq:guiding}) provides a qualitative approximation of the electron fluid energization as quantified by $\boldsymbol{J}_e\cdot\boldsymbol{E}$ when summed over the whole simulation domain (differences up to around $20\%$), it fails to reproduce $\boldsymbol{J}_e\cdot\boldsymbol{E}$ on a local, point-by-point, level. The authors then derived a complete Eulerian fluid decomposition of $\boldsymbol{J}_e\cdot\boldsymbol{E}$, and expressed it in several different ways, one of which is
\begin{align}
\begin{split}
\label{eq:fluid}
    \boldsymbol{J}_e\cdot\boldsymbol{E} &= J_{e\parallel}E_\parallel + \boldsymbol{u}_E\cdot\nabla p_{e\perp}+(p_{e\parallel}-p_{e\perp})\boldsymbol{u}_E\cdot\boldsymbol{k}+\boldsymbol{u}_E\cdot\left(\nabla\cdot\boldsymbol{\Uppi}_e^a\right)+m_en_e\boldsymbol{u}_E\cdot\frac{d\boldsymbol{u}_e}{dt} \\
    &:= W_\parallel + W_\text{diam} + W_\text{curv1} + W_\text{agyro} + W_\text{pol},
\end{split}
\end{align}
where $\boldsymbol{\Uppi}^a_e = \mathbf{P}_e - p_{e\perp}\mathbf{I} - (p_{e\parallel} - p_{e\perp}) \boldsymbol{b}\boldsymbol{b}$ is the (traceless) agyrotropic part of the electron pressure tensor defined as $\mathbf{P}_e=m_e\int d^3v (\boldsymbol{v}-\boldsymbol{u}_e)(\boldsymbol{v}-\boldsymbol{u}_e) f_e$, and $\mathbf{I}$ is the identity matrix. $W_\text{diam}$ is related to the diamagnetic drift, $W_\text{curv1}$, to the Eulerian curvature drift,  $W_\text{agyro}$ to the agyrotropic drift, and $W_\text{pol}$ to the polarisation drift. We will refer to this model as the Eulerian fluid model (EFM). Eq.\ (\ref{eq:fluid}) was derived without approximations, and it was found to accurately describe $\boldsymbol{J}_e\cdot\boldsymbol{E}$ on a global and local level. For the case studied by \citet{tenbarge2024}, the authors found that the energization term related to the diamagnetic drift ($W_\text{diam}$) was, by far, the most dominant on system scales, while all terms could have significant local contributions to $\boldsymbol{J}_e\cdot\boldsymbol{E}$. We note in passing that other Eulerian decompositions of $\boldsymbol{J}_e\cdot\boldsymbol{E}$, corresponding to different groupings of the different terms, have previously been studied, such as the fluid compression/shear decomposition by \citet{li2018}. However, in that specific case the electrons are assumed to be well magnetized such that the agyrotropic contribution is neglected.

The \citet{tenbarge2024} study raises several interesting questions. Firstly, as previously mentioned, it is well known that magnetic structures such as magnetic islands play an important role in accelerating electrons to high energies through Fermi-like processes \citep{drake2006}. Such islands are formed when reconnection is triggered at multiple locations along the current sheet \citep[e.g.][]{karimabadi2005,drake2006b}. Therefore, it is of interest to study the EFM also in the multiple x-point case, and compare it to the LGCM. On a related note, we know that the relative importance of the LGCM terms depends on the guide field strength \citep{dahlin2016}, so it is natural to ask how the EFM terms depend on the guide field. We thus pose the question: which of the EFM terms are most important when reconnection leads to the formation of magnetic islands, and is there a guide field dependence? 

Secondly, the inability of the LGCM to reproduce $\boldsymbol{J}_e\cdot\boldsymbol{E}$ begs the question what is the reason for the erroneous LGCM prediction? Is it related to a false assumption of $\mu$-conservation, is it due to the Eulerian/Lagrangian framework inconsistency discussed by \citet{tenbarge2024}, or is it something else? Related to this, what happens to the electrons in the regions where the LGCM and EFM predict significantly different $\boldsymbol{J}_e\cdot\boldsymbol{E}$? 

Finally, one important feature of the LGCM is that each $W$ term can be interpreted as having a different effect on the electron velocity distribution function. The betatron term ($W_\text{beta-gradB}$) is associated with perpendicular energization, while the curvature (or Fermi) term ($W_\text{curv0}$) is associated with parallel energization and ultimately the formation of power-law tails if repeated interactions are possible. There is a pedagogical value in these types of simple interpretations as they can aid in developing physical intuition. With this in mind, it is tempting to ask the question: can we make analogous interpretations of the fluid terms in Eq.\ (\ref{eq:fluid})? That is, can we associate local measurements of each EFM $W$-term with the generation of specific VDF features?
 
In the present paper, we answer the aforementioned questions using particle-in-cell (PIC) simulations of multi-x-point reconnection. We find that $W_\text{diam}$ is the dominant EFM contributor to $\boldsymbol{J}_e\cdot\boldsymbol{E}$ on large scales for low and intermediate guide fields, $B_g\in\{0,0.2\}$, where $B_g$ is normalized to the reconnecting field component. For large guide fields ($B_g = 1$), $W_\parallel$ instead becomes the largest contributor (albeit still comparable to $W_\text{diam}$). Locally, $W_\text{agyro}$ and $W_\text{curv1}$ can contribute significantly to electron energization, particularly in the lower $B_g$ regime. We show that the assumption of $\mu$-conservation made by the LGCM is invalid in the low $B_g$ case near the x-points and at the center of the current sheets. This causes significant errors in the LGCM energization estimate. Moreover, we find that local deviations from $\boldsymbol{J}_e\cdot\boldsymbol{E}$ are produced by the LGCM even when $\mu$ is conserved. This is due to the fact that the LGCM, unlike the EFM, does not describe the energy flux through the grid. Finally, we argue that we cannot easily associate local measurements of the different EFM energization terms with the formation of features in the electron VDF.

\section{Numerical Setup}
\label{sec:numerical_setup}
In this study, we use the OSIRIS PIC-code \citep{fonseca2002} to study electron energization during multi-x-point magnetic reconnection in 2 spatial and 3 velocity dimensions. We use a Harris current sheet \citep{harris1962} setup with periodic boundary conditions in $x$, and perfectly conducting boundaries in $y$. We let the reconnection process grow from numerical noise.
For all runs, we use an unperturbed magnetic field profile $B_x(y)=B_\infty\tanh{(y/\lambda)}$ with an asymptotic magnetic field strength $B_\infty$ corresponding to $\omega_{ce\infty}/\omega_{pe0}=1/3$. Here, the electron cyclotron frequency $\omega_{ce}=|q_e| B/m_e$ is evaluated at $B=B_\infty$, and $\omega_{pe0}=\sqrt{q_e^2 n_0/(m_e\epsilon_0)}$ is the reference electron plasma frequency, using the Harris sheet density perturbation $n_0$. The complete density profile is $n_e=n_0\text{sech}^2(y/\lambda)+n_\infty$, where we use $n_\infty/n_0=0.2$, and a current sheet thickness of $\lambda/d_{e0}=1.25$, where $d_{e0}=c/\omega_{pe0}$ is the electron inertial length. The narrow current sheet thickness was chosen to be consistent with \citet{dahlin2014}, and for reconnection to trigger quickly. We use two plasma components, one electron population, and one ion population, with a uniform ion-to-electron temperature ratio of $T_i/T_e=1$.

As our baseline numerical resolution, we use a time step $0.15\omega_{pe0}^{-1}$ and cell size $\Delta x=\Delta y=0.25d_{e0}$. The size of the simulation domain is $L_x\times L_y=51.2d_{i0}\times12.8d_{i0}$, where $d_{i0}=d_{e0}\sqrt{m_i/m_e}$ is the ion inertial length. Each cell is initialized with 625 macroparticles of each species. We use guide field values of $B_g/B_{\infty}\in\{0,0.2,1\}$, and low mass ratios $m_i/m_e\in\{25,100\}$ similar to \citet{dahlin2014} and \citet{tenbarge2024}. We note that neither of the aforementioned studies investigated the $B_g=0$ case, to avoid the problem of demagnetized electrons breaking the LGCM assumption of $\mu$ conservation. 

%The simulation parameters are summarized in table \ref{tab:simulations}.
%\begin{table}
%  \begin{center}
%\def~{\hphantom{0}}
%  \begin{tabular}{lccc}
%       Run ID & $B_g$  & $m_i/m_e$ \\[3pt]
%       1    & 0     & 25\\
%       2    & 0.2   & 25\\
%       3    & 1     & 25\\
%       4    & 0     & 100\\
%       5    & 0.2   & 100\\
%  \end{tabular}
%  \caption{Simulation parameters}
%  \label{tab:simulations}
%  \end{center}
%\end{table}

Henceforth, unless otherwise specified (e.g. in figure axes with explicit normalization), we use the following normalizations in our calculations. Time is normalized by $\omega_{pe0}^{-1}$, lengths by $d_{e0}$, $\boldsymbol{E}$ and $c\boldsymbol{B}$ by $m_ec^2/(ed_{e0})$, where $c$ is the speed of light and $e$ is the elementary charge. Compound quantities such as $\boldsymbol{J}_e\cdot\boldsymbol{E}$ are normalized accordingly.

\section{Results}
\label{sec:results}
\subsection{Global energization and parametric dependence}
\label{sec:global}
\begin{figure}
    \centering
    \includegraphics[width=1\linewidth]{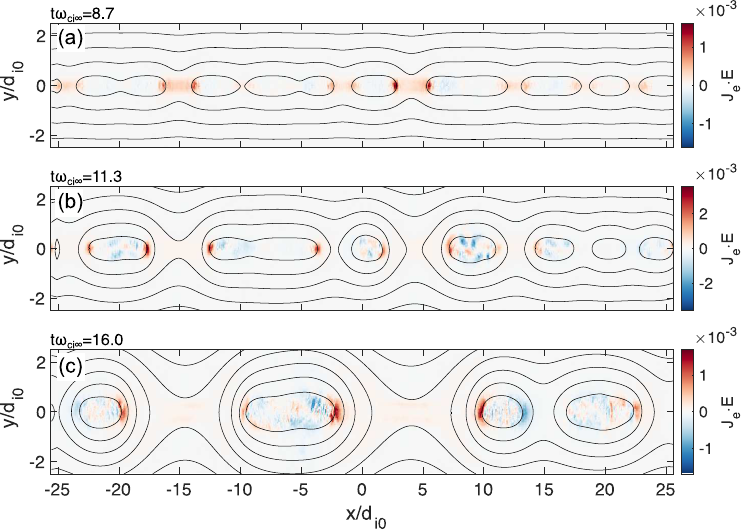}
    \caption{Snapshots of $\boldsymbol{J}_e\cdot\boldsymbol{E}$ for the $m_i/m_e=25$, $B_g=0$ run at three different times $t\omega_{ci\infty}\in\{8.7,11.3,16.0\}$, where $\omega_{ci\infty}=\omega_{ce\infty}m_e/m_i$. The black lines are contours of the vector potential component $A_z$. \label{fig:evolution}}
\end{figure}

\begin{figure}
    \centering
    \includegraphics[width=\linewidth]{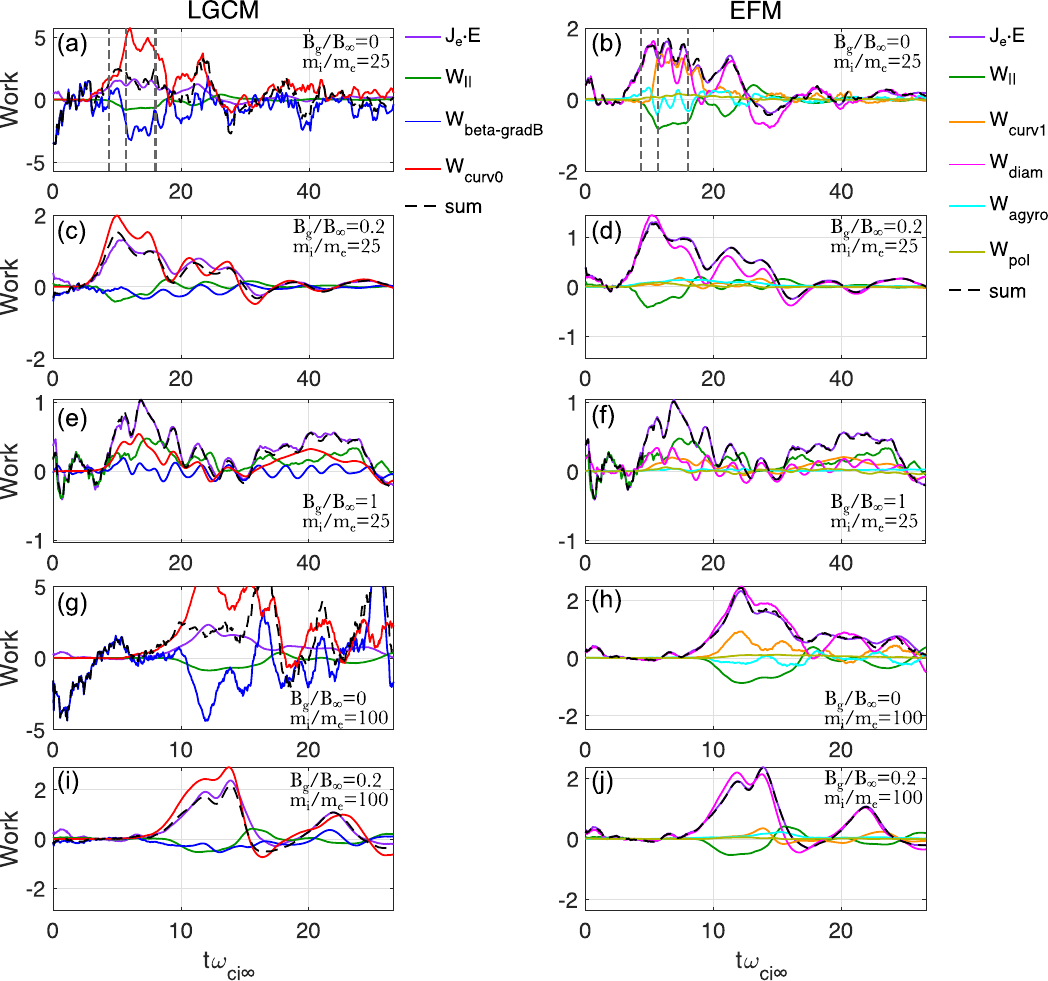} 
    \caption{Time evolution of energization terms summed over the whole domain (arbitrary units) for different $m_i/m_e\in\{25,100\}$ and $B_g/B_\infty\in\{0,0.2,1\}$ as labeled in the panels. The left column shows the guiding center model results from Eq.~(\ref{eq:guiding}) and the right column the corresponding fluid model results from Eq.~(\ref{eq:fluid}). The three vertical dashed lines in panels (a) and (b) correspond to the times of the panels in Figure \ref{fig:evolution}. Note that we plot a shorter time interval for the $m_i/m_e=100$ simulations in panels (g-j). The y-axis of panel (g) has been limited to better show the deviation between $\boldsymbol{J}_e\cdot\boldsymbol{E}$ and the LGCM sum. All data have been smoothed using a moving average with a window of $\pm 1\omega_{ci\infty}^{-1}$ to reduce noise.}
    \label{fig:w_time}
\end{figure}

We will start by investigating the energization models on large scales.
The temporal evolution of $\boldsymbol{J}_e\cdot\boldsymbol{E}$ is shown in Figure \ref{fig:evolution}, for the $m_i/m_e=25$, $B_g=0$ run. The evolution is qualitatively similar for all runs. As reconnection starts and magnetic islands start to form (Figure \ref{fig:evolution}a), patches of positive $\boldsymbol{J}_e\cdot\boldsymbol{E}$ are found at the x-points and the edges of the magnetic islands. The total contribution of the different energization terms in Eqs. (\ref{eq:guiding}) and (\ref{eq:fluid}) over the whole simulation domain for this time are shown in Figures~\ref{fig:w_time}a and \ref{fig:w_time}b, marked by the first vertical line. We find that the curvature/Fermi term $W_\text{curv0}$ is the dominant contributor to $\boldsymbol{J}_e\cdot\boldsymbol{E}$ in the LGCM, while the diamagnetic term $W_\text{diam}$ dominates in the EFM. The dominance of these two terms is consistent with the previous studies by \citet{dahlin2014} and \citet{tenbarge2024}.

As the system develops further (Figure \ref{fig:evolution}b), the islands contract, coalesce, and move with the embedding plasma flow. Locally, $\boldsymbol{J}_e\cdot\boldsymbol{E}$ varies depending on the local island dynamics. The largest values of $\boldsymbol{J}_e\cdot\boldsymbol{E}$ are found in narrow regions at the magnetic island edges. These regions are associated with local contraction of the island, and Fermi-like processes are expected to occur there \citep{drake2006,dahlin2014}. Similar regions of negative $\boldsymbol{J}_e\cdot\boldsymbol{E}$ corresponds to local expansion of the island. Consequently, islands that are convecting have a bipolar $\boldsymbol{J}_e\cdot\boldsymbol{E}$ signature, where the leading edge (locally expanding) is negative, and the trailing edge (locally contracting) is positive. At $t\omega_{ci\infty}=11.3$ (second vertical lines in Figure \ref{fig:w_time}a and \ref{fig:w_time}b), the LGCM finds that the Fermi related $W_\text{curv0}$ is indeed large, while $W_\parallel$ gives a slight negative contribution to $\boldsymbol{J}_e\cdot\boldsymbol{E}$ together with $W_\text{beta-gradB}$. In the case of the EFM, $W_\text{diam}$ is still large, but now there are also significant net contributions from $W_\text{curv1}>0$, $W_\parallel<0$, and $W_\text{agyro}<0$, showing that multiple fluid drifts contribute to the net energization. 
The same general picture remains valid as the islands continue to grow and merge (Figure \ref{fig:evolution}c).

Looking at the temporal evolution of the $W$-terms in Figures \ref{fig:w_time}a and \ref{fig:w_time}b, we find that the LGCM generally provides a qualitative approximation of $\boldsymbol{J}_e\cdot\boldsymbol{E}$ (compare the purple and the dashed black curves), particularly during the time interval $t\omega_{ci\infty}\in(7,18)$ when $\boldsymbol{J}_e\cdot\boldsymbol{E}$ is large. However, at times the relative difference between the LGCM sum and $\boldsymbol{J}_e\cdot\boldsymbol{E}$ exceeds $80\%$. Integrating over the duration of the simulation reduces the relative difference to around $10\%$.
In the EFM case, the agreement between the sum of the $W$-terms and $\boldsymbol{J}_e\cdot\boldsymbol{E}$ is almost perfect. These results are consistent with those from \citet{tenbarge2024} in the single x-point case.

The remaining panels in Figure \ref{fig:w_time} show the results of the same analysis applied to four other simulations with varying mass-ratio and guide field. Figures \ref{fig:w_time}c and \ref{fig:w_time}d show the case with moderate guide field $B_g=0.2$, at the same mass ratio  $m_i/m_e=25$. There, the LGCM performs somewhat better than in the $B_g=0$, $m_i/m_e=25$ case discussed previously, and the betatron $W_\text{beta-gradB}$ term is slightly less important than before. In the EFM case, we find that the increased $B_g$ almost removes the curvature contribution entirely, and the agyrotropic contribution also becomes less important. This is expected since including an out-of-plane magnetic field component reduces the magnetic curvature across the current sheet and, consequently, the curvature drift. In addition, the guide field contributes to keeping the electrons magnetized \citep[e.g.][]{swisdak2005,scudder2015}, reducing the agyrotropic pressure carried by demagnetized electrons. Increasing the guide field to $B_g=1$ (Figures \ref{fig:w_time}e and \ref{fig:w_time}f) further improves the accuracy of the LGCM to a very good agreement with $\boldsymbol{J}_e\cdot\boldsymbol{E}$. The relative contribution of $W_\text{curv0}$ decreases, while $W_\parallel$, which for lower $B_g$ was a negative contributor to $\boldsymbol{J}_e\cdot\boldsymbol{E}$, becomes large and positive. Analogously, in the EFM (Figure \ref{fig:w_time}f) $W_\text{diam}$ decreases and $W_\parallel$ becomes the largest energization term, albeit still comparable to $W_\text{diam}$. The increased importance of $W_\parallel$ in the LGCM for increasing guide fields was previously reported by \citet{dahlin2016}.

Increasing the mass ratio to $m_i/m_e=100$ does not change the results significantly. In the $B_g=0$ case, the LGCM results (Figure~\ref{fig:w_time}g) are qualitatively very similar to the $m_i/m_e=25$ case (Figure~\ref{fig:w_time}a), with $W_\text{curv0}$ being the dominant positive contributor, while $W_\parallel$ and $W_\text{beta-gradB}$ are negative. In the EFM case (Figure~\ref{fig:w_time}h) the relative contribution of $W_\text{curv1}$ is noticeably smaller than in the $m_i/m_e=25$ case, while the other terms are less affected by the $m_i/m_e$ change. Finally, in the $B_g=0.2$ case (Figures \ref{fig:w_time}i and \ref{fig:w_time}j), the increased mass ratio does not affect the relative importance of terms in either of the models (cf. Figures \ref{fig:w_time}c and \ref{fig:w_time}d). 

In summary, we conclude that the domain-integrated LGCM is generally dominated by $W_\text{curv0}$ and the EFM is generally dominated by $W_\text{diam}$ when $B_g\leq0.2$. Increasing the guide field to $B_g=1$ leads to $W_\parallel$ becoming larger than (or comparable to) $W_\text{curv0}$ and $W_\text{diam}$. The only significant $m_i/m_e$ dependence is found for the Eulerian $W_\text{curv1}$, which becomes less important in the $B_g=0$ case as $m_i/m_e$ is increased from 25 to 100. While the EFM accurately reproduces the total $\boldsymbol{J}_e\cdot\boldsymbol{E}$ over the simulation domain for all runs, we find that the LGCM estimate of electron energization agrees better with $\boldsymbol{J}_e\cdot\boldsymbol{E}$ for increasing values of $B_g$. Since the LGCM assumes $\mu$-conservation, and a larger $B_g$ leads to stronger magnetization, this result hints that the deviation between the LGCM and $\boldsymbol{J}_e\cdot\boldsymbol{E}$ is related to the presence of non-adiabatic electrons.

Similarly to \citet{tenbarge2024}, we find that the local and volume-integrated pictures of electron energization are very different. Next, we will investigate the spatial distribution of $\boldsymbol{J}_e\cdot\boldsymbol{E}$ and the energization terms for the $m_i/m_e=25$ case in more detail.

\subsection{Local differences between the LGCM and EFM descriptions}
\label{sec:local}
\begin{figure}
    \centering
    \includegraphics[width=1\linewidth]{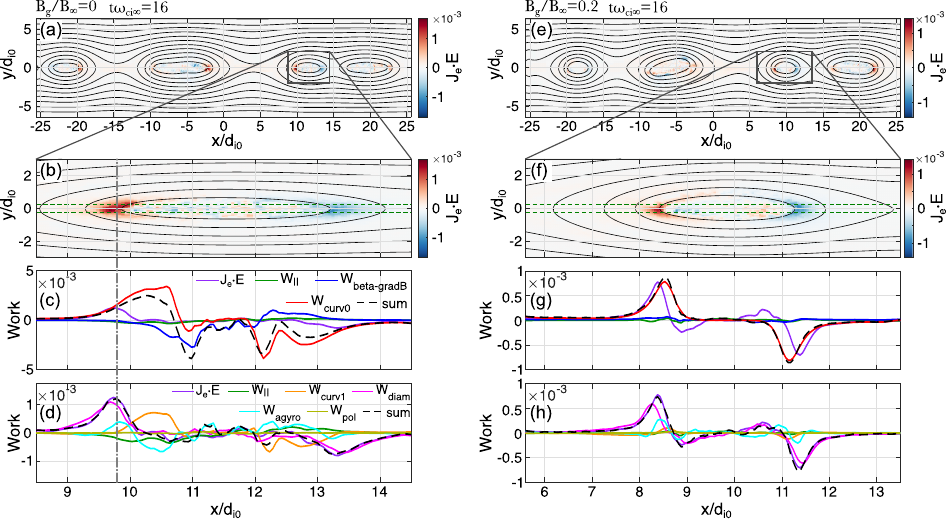}
    \caption{Spatial dependence of $\boldsymbol{J}_e\cdot\boldsymbol{E}$ and the $W$ terms inside magnetic islands for $B_g=0$ (left column) and $B_g=0.2$ (right column). (a) $\boldsymbol{J}_e\cdot\boldsymbol{E}$ for the whole simulation domain. (b) $\boldsymbol{J}_e\cdot\boldsymbol{E}$ for the magnetic island boxed in panel (a). The black lines are contours of $A_z$. The dashed green lines show the y-range in which the data in panels (c) and (d) are averaged. (c) The different terms in the guiding center model (Eq.~\ref{eq:guiding}) as a function of $x$. (d) Same as (c) for the fluid model (Eq.~\ref{eq:fluid}). The vertical dash-dotted line indicates where the LGCM and $\boldsymbol{J}_e\cdot\boldsymbol{E}$ start to qualitatively deviate.
    The data in panels (c) and (d) have been averaged in $y$ over the interval marked by the dashed green lines in panel (b), and smoothed in $x$ by a moving mean of window size $\pm0.5d_{e0}$. (f-h) Same format as (a-d) for the $B_g=0.2$ case.}
    \label{fig:local_energization}
\end{figure}

The spatial structures of $\boldsymbol{J}_e\cdot\boldsymbol{E}$ and the $W$-terms of Eqs.~(\ref{eq:guiding}) and (\ref{eq:fluid}) are shown in Figure~\ref{fig:local_energization}. The left column shows the results of the $B_g=0$ simulation at $t\omega_{ci\infty}=16$. As was seen in Figure \ref{fig:evolution}, we find the strongest $\boldsymbol{J}_e\cdot\boldsymbol{E}$ at the edges of the magnetic islands (Figures \ref{fig:local_energization}a and \ref{fig:local_energization}b), and we will focus our attention to these regions.

In the LGCM, $W_\text{curv0}$ (Figure \ref{fig:local_energization}c, red) is the dominant energization term due to the strong magnetic curvature associated with the island. In the Eulerian view, the strong diamagnetic drift associated with the pressure gradient at the island edges results in significant $W_\text{diam}$ (Figure \ref{fig:local_energization}d, magenta), which is the main contributor to $\boldsymbol{J}_e\cdot\boldsymbol{E}$. Approaching the island from the left, the sum of the energization terms in Eq.~(\ref{eq:guiding}) and Eq.~(\ref{eq:fluid}) (black dashed lines in Figures \ref{fig:local_energization}c and \ref{fig:local_energization}d) are both in good agreement with $\boldsymbol{J}_e\cdot\boldsymbol{E}$ (purple) until around $x/d_{i0}\approx9.8$, marked by the dash-dotted vertical line. After this point, $\boldsymbol{J}_e\cdot\boldsymbol{E}$ starts to decrease. This is well captured by the EFM, which reveals a decreasing $W_\text{diam}$ while $W_\text{curv1}\approx -W_\text{agyro}$. In contrast, the LGCM deviates from $\boldsymbol{J}_e\cdot\boldsymbol{E}$ by predicting further energization deeper into the island as a result of a large $W_\text{curv0}$. Analogous results (with opposite signs) are obtained for the right edge of the island where $\boldsymbol{J}_e\cdot\boldsymbol{E}$ has a negative peak.

In the $B_g=0.2$ case (right column of Figure \ref{fig:local_energization}), the LGCM works better than in the $B_g=0$ case, but it still predicts energization slightly deeper into the island than the measured $\boldsymbol{J}_e\cdot\boldsymbol{E}$. In both cases, $W_\text{curv0}$ is the dominant term. The EFM remains accurate, with the main difference being that $W_\text{curv1}$ is less important than in the $B_g=0$ case. The local differences between the models brings us to the second question posed in the introduction: why does the LGCM fail inside the islands, and what actually happens to the electrons at these locations?

Since the LGCM assumes adiabatic electrons, we start by scrutinizing this assumption. One important parameter which can be used to quantify whether the motion of a particle is adiabatic is the so-called adiabaticity parameter $\kappa=\sqrt{r_\text{curv}/r_{ge}}$, where $r_\text{curv}=1/|\boldsymbol{k}|$ is the magnetic curvature radius and $r_{ge}=v_{\perp} m_e/(eB)$ the electron gyroradius \citep{buchner1989}. An electron with $\kappa<1$ experiences strong magnetic field changes over its gyro-orbit, and its magnetic moment is not conserved. We note that one may define a generalized adiabaticity parameter, accounting for all spatial variations of $\mathbf{B}$. We do that in Appendix~\ref{gradBmeasure} for reference, but since the alternative definition does not affect the following discussion qualitatively, we will continue using the more widely used definition, given above.

\begin{figure}
    \centering
    \includegraphics[width=1\linewidth]{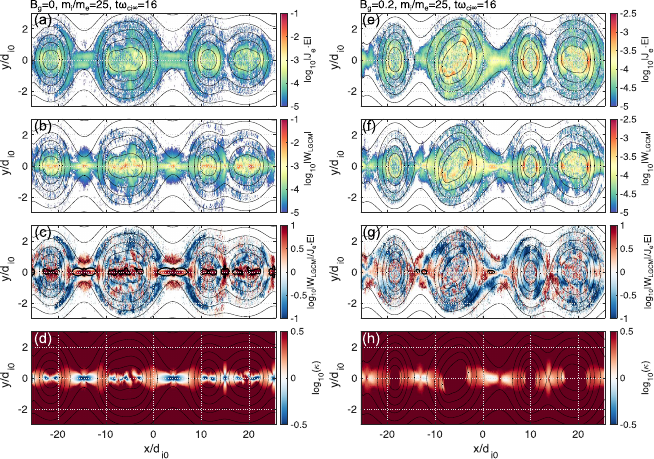}
    \caption{Deviations from $\boldsymbol{J}_e\cdot\boldsymbol{E}$ in the LGCM for $B_g=0$ (a-d) and $B_g=0.2$ (e-h). Panels (a) and (b) show respectively $\boldsymbol{J}_e\cdot\boldsymbol{E}$ and the sum of the guiding center model terms in Eq.~(\ref{eq:guiding}), $W_\text{LGCM}$. Only values larger than $10^{-5}$ have been included. (c) $|W_\text{LGCM}/\boldsymbol{J}_e\cdot\boldsymbol{E}|$ using the data in (a) and (b). The thick black contour shows the $\kappa=1$ threshold, where the bounded regions contain $\kappa<1$. The colorbar is saturated. (d) The adiabaticity parameter $\kappa$ computed using the perpendicular thermal speed. Blue regions correspond do $\kappa<1$, i.e.\ where the motion of thermal electrons is not adiabatic, and red regions to $\kappa>1$. The colorbar has been limited to the range $\log_{10}(\kappa)\in[-0.5,0.5]$ to highlight the adiabatic/non-adiabatic transition. (e-h) Same format as (a-d) for the $B_g=0.2$ case.}
    \label{fig:guiding_error}
\end{figure}

In Figures \ref{fig:guiding_error}a, \ref{fig:guiding_error}b, and \ref{fig:guiding_error}c, we present $|\boldsymbol{J}_e\cdot\boldsymbol{E}|$, the sum of the right hand side of Eq.~(\ref{eq:guiding}), $|W_\text{LGCM}|$, and $|W_\text{LGCM}/\boldsymbol{J}_e\cdot\boldsymbol{E}|$ for the $B_g=0$ case. The largest relative discrepancies between the LGCM and $\boldsymbol{J}_e\cdot\boldsymbol{E}$ are found at the x-points and in the center of the magnetic islands (Figure \ref{fig:guiding_error}c), where $|W_\text{LGCM}|$ is up to a few orders of magnitude larger than $|\boldsymbol{J}_e\cdot\boldsymbol{E}|$. These locations are regions where the thermal $\kappa$ (using $v_{te \perp}=\sqrt{2T_{e\perp}/m_e}$ in $r_{ge}$) is very small ($\kappa\ll1$), as shown in Figure~\ref{fig:guiding_error}d. Since $\kappa\ll1$ implies non-adiabatic electron motion, we conclude that the deviations in these regions can likely be attributed to the erroneous assumption of $\mu$ conservation made by the LGCM. The fact that electrons can exhibit non-adiabatic behavior during magnetic reconnection is well known \citep{zenitani2016}, and the failure of the LGCM near the current sheet center and at the x-points in the low $B_g$ limit is not surprising.
In other regions, the error does not correlate well with $\kappa$. One such example is the separatrices of the larger reconnection sites. There, the LGCM deviates from $\boldsymbol{J}_e\cdot\boldsymbol{E}$ even though $\kappa\gg1$.

In the $B_g=0.2$ case, the presence of the out-of-plane component leads to a reduced magnetic curvature (i.e.\ a larger $r_\text{curv}$), and we generally find $\kappa\gg1$ (Figure~\ref{fig:guiding_error}h). Small values of $\kappa$ are still found near the x-points, and $\kappa$ occasionally drops below 1. It should be noted that the plotted $\kappa$ is for thermal electrons, and electrons with $v_\perp>v_{te\perp}$ have smaller $\kappa$. A large fraction of the electrons might therefore be non-adiabatic when the thermal $\kappa\approx1$. The fact that there are still significant deviations even though $\kappa>1$ (Figures \ref{fig:local_energization}g and \ref{fig:guiding_error}h) suggests that there is another issue at play other than the lack of $\mu$-conservation. This suspicion is reinforced by the fact that the LGCM locally deviates from $\boldsymbol{J}_e\cdot\boldsymbol{E}$ even in the $B_g=1$ case (see Appendix \ref{app:bg1}), where $\kappa\gg1$ everywhere. 

We conclude that the non-adiabatic motion of electrons near x-points and the current sheet center, particularly in the $B_g=0$ case, is a large source of error in the LGCM. However, this is not the entire picture, and there is still the issue of the transition between the Lagrangian and Eulerian frameworks when Eq.~(\ref{eq:guiding}) is evaluated on an Eulerian grid. This problem was discussed by \citet{tenbarge2024}, and in the following section, we will expand on their discussion to pin-point the issue and illustrate how this affects the local electron energization estimate.

\subsection{Eulerian vs Lagrangian energization}
\label{sec:eulerian_vs_lagrangian}
Specifically, the problem with evaluating Eq.\ (\ref{eq:guiding}) on an Eulerian grid (assuming $\mu$ is conserved) is that it gives the impression that it provides an Eulerian view on electron energization, i.e., that it describes the evolution of the total electron energy within a fixed volume element. 
However, Eq.\ (\ref{eq:guiding}) is derived in a Lagrangian framework from the single guiding center equation (Eq.~(\ref{eq:single_guiding})), which describes the rate of energy gain a single guiding center experiences at some given time. To derive Eq.\ (\ref{eq:guiding}) from Eq.\ (\ref{eq:single_guiding}), one takes a small volume (such that all charges experience the same fields) and adds up Eq.\ (\ref{eq:single_guiding}) for all guiding centers within the volume. Thus, what the quantity $d\mathcal{E}_\text{LGCM}/dt$ in Eq.\ (\ref{eq:guiding}) describes is the rate of energy gain per unit volume of the specific guiding centers that are currently in the volume element. 
Importantly, this means that the electron energy flux into or out of the volume is irrelevant for $d\mathcal{E}_\text{LGCM}/dt$. This detail distinguishes the Lagrangian description from the Eulerian description, which is concerned with the evolution of the total electron energy content within the volume. The Eulerian energy equation therefore contains divergence terms related to the electron energy flux:
\begin{equation}
\label{eq:energy_eq}
    \frac{\partial\mathcal{E}_\text{EFM}}{\partial t} + \nabla\cdot\left(\boldsymbol{K}+\boldsymbol{H}+\boldsymbol{q}\right) = \boldsymbol{J}_e\cdot\boldsymbol{E},
\end{equation}
where $\mathcal{E}_\text{EFM}=m_en_eu_e^2/2+\text{trace}(\mathbf{P}_e)/2:=K+U$ is the total electron energy density with $K$ and $U$ being respectively the bulk kinetic and thermal contributions, $\boldsymbol{K}=K\boldsymbol{u}_e$ is the kinetic energy flux density, $\boldsymbol{H}=U\boldsymbol{u}_e+\mathbf{P}_e\cdot\boldsymbol{u}_e$ is the enthalpy flux density, and $\boldsymbol{q}$ is the heat flux density \citep[e.g.][]{helander2005}. The energy equation describes the evolution of the electron energy density inside a fixed volume element, a quantity that is fundamentally different from the aforementioned Lagrangian $d\mathcal{E}_\text{LGCM}/dt$.

As evident from Eq.\ (\ref{eq:energy_eq}), $\boldsymbol{J}_e\cdot\boldsymbol{E}$ is a source term and it can be balanced by the time derivative and divergence (i.e.\ energy flux density) terms.
\begin{figure}
    \centering
    \includegraphics[width=1\linewidth]{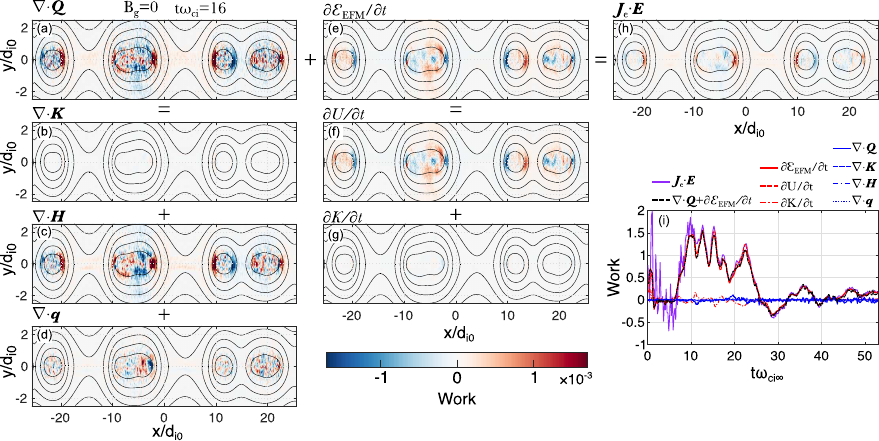}
    \caption{The contribution of the different terms in Eq.\ (\ref{eq:energy_eq}) to $\boldsymbol{J}_e\cdot\boldsymbol{E}$ in the $B_g=0$, $m_i/m_e=25$ simulation.
    (a-h) Spatial profiles for the different terms at $t\omega_{ci\infty}=16$. The top row of panels show the three terms in Eq.~(\ref{eq:energy_eq}), where $\boldsymbol{Q}=\boldsymbol{K}+\boldsymbol{H}+\boldsymbol{q}$. The first and second columns of panels show the different divergence and time derivative terms, respectively. The divergence terms (a-d) have been smoothed using a 5-point moving average to reduce noise. (i) Time dependence of the different terms summed over the whole simulation domain.}
    \label{fig:eflux}
\end{figure}
The importance of the divergence terms on smaller scales is readily seen in Figure \ref{fig:eflux}, where the spatial distribution of the different components of Eq.\ (\ref{eq:energy_eq}) are shown in panels a-h.
The large $\boldsymbol{J}_e\cdot\boldsymbol{E}$ found at the edges of magnetic islands is mainly balanced by the divergence terms in the energy equation (see Figure \ref{fig:eflux}a and \ref{fig:eflux}h, where $\boldsymbol{Q}=\boldsymbol{K}+\boldsymbol{H}+\boldsymbol{q}$), where $\nabla\cdot\boldsymbol{H}$ dominates (Figure~\ref{fig:eflux}c) over $\nabla\cdot\boldsymbol{K}$ (Figure~\ref{fig:eflux}b) and $\nabla\cdot\boldsymbol{q}$ (Figure~\ref{fig:eflux}d). The time derivative terms (Figure~\ref{fig:eflux} e-g) tend to have the opposite sign to the divergence terms, and the internal energy term (Figure~\ref{fig:eflux}f) dominates over the kinetic energy term (Figure~\ref{fig:eflux}g). The large contribution of the energy flux terms, particularly $\nabla\cdot\boldsymbol{H}$, is consistent with in-situ studies of magnetic reconnection \citep{eastwood2020,fargette2024}.
We conclude that, on a local level, most of $\boldsymbol{J}_e\cdot\boldsymbol{E}$ is balanced by the electron energy flux densities, with the dominant term being the enthalpy flux density. This is a key reason why the LGCM locally deviates from $\boldsymbol{J}_e\cdot\boldsymbol{E}$, even when the assumption of magnetic moment conservation is valid.

On system scales, however, the picture is very different. For our choice of boundary conditions, the system is closed in the sense that there is no net particle energy flux into or out of the simulation domain. Therefore, when we integrate the divergence terms of Eq.\ (\ref{eq:energy_eq}) over the whole simulation, the corresponding surface terms vanish, leaving us with $\boldsymbol{J}_e\cdot\boldsymbol{E}\approx\partial\mathcal{E}_\text{EFM}/\partial t$, as seen in Figure \ref{fig:eflux}i. This is why the domain-integrated LGCM can accurately reproduce $\boldsymbol{J}_e\cdot\boldsymbol{E}$ when $\mu$-conservation applies (Figure \ref{fig:w_time}e). In that case, Eq.\ (\ref{eq:guiding}) describes the evolution of the total energy of all guiding centers, and since they are confined to the simulation domain, this is the same thing as the evolution of the total electron energy in the simulation domain. So, when we look at a closed system in its entirety, indeed $\int d^3x(d\mathcal{E}_\text{LGCM}/dt) = \int d^3x(\partial\mathcal{E}_\text{EFM}/\partial t)=\int d^3x\boldsymbol{J}_e\cdot\boldsymbol{E}$. The difference between the domain-integrated LGCM and $\boldsymbol{J}_e\cdot\boldsymbol{E}$ observed in the $B_g=0$ case (Figure \ref{fig:w_time}a) can thus be attributed to the lack of $\mu$ conservation due to the small $\kappa$ (Figure \ref{fig:guiding_error}d). The fact that the LGCM equation is locally invalid in the $B_g=0$ case explains why the domain-integrated LGCM both overestimates and underestimates $\boldsymbol{J}_e\cdot\boldsymbol{E}$ in the $B_g=0$ case (Figure \ref{fig:w_time}a), and subsequently why the time-integrated energization more accurately reproduces $\boldsymbol{J}_e\cdot\boldsymbol{E}$.  

Until this point, we have followed \citet{tenbarge2024} and uncritically assumed that, since $d\mathcal{E}_\text{LGCM}/dt$ accurately describes the energization of individual guiding centers, it must equal the work done on the electrons as quantified by $\boldsymbol{J}_e\cdot\boldsymbol{E}$. In other words, since we can write Eq.\ (\ref{eq:guiding}) as the scalar product of a net current density $\boldsymbol{J}_{e,\text{LGCM}}$ and $\boldsymbol{E}$, $\boldsymbol{J}_{e,\text{LGCM}}$ should be equal to the bulk fluid current density $\boldsymbol{J}_e$. This, however, is evidently not the case. As already mentioned by \citet{tenbarge2024}, the guiding center model is a particle model that is gyroperiod integrated, and it consequently does not describe certain bulk fluid properties such as the diamagnetic drift and the corresponding currents. Therefore, in general $\boldsymbol{J}_{e,\text{LGCM}}\neq\boldsymbol{J}_e$, since $\boldsymbol{J}_e$ describes the total electron current. In hindsight, it is therefore quite clear that we should never have expected $d\mathcal{E}_\text{LGCM}/dt$ to accurately model $\boldsymbol{J}_e\cdot\boldsymbol{E}$ to begin with, even when $\mu$ is conserved.

Finally, we note that while Eq.\ (\ref{eq:guiding}) is derived in the Lagrangian framework, the right-hand-terms are expressed in terms of bulk fluid quantities which, in our case, are known on the points defined by the Eulerian grid. It is perfectly valid to evaluate Eq.\ (\ref{eq:guiding}) on an Eulerian grid (if $\mu$ is conserved); one must just keep in mind that a local measurement of $d\mathcal{E}_\text{LGCM}/dt$ describes the energy evolution of the guiding centers that are presently in the local volume, not the evolution of the electron energy within the volume. In their paper, \citet{tenbarge2024} state, regarding the LGCM results, that \enquote{[...] neither globally nor locally does the sum of the energization mechanisms agree well with $\boldsymbol{J}_e\cdot\boldsymbol{E}$, suggesting that the Lagrangian description, unsurprisingly, does not work well in an Eulerian simulation}, which may give the impression that there is an inherent conflict between their Eulerian solver and the LGCM. We want to stress that the inherently Eulerian nature of a continuum Vlasov-Maxwell solver does not affect the applicability of Eq.\ (\ref{eq:guiding}). The deviations from $\boldsymbol{J}_e\cdot\boldsymbol{E}$ observed by \citet{tenbarge2024} on a global level can likely be attributed to the demagnetization of electrons near the x-point, as their choice of $B_g=0.1$ is not large enough to keep the electrons well magnetized (see our $B_g=0.2$ and $B_g=1$ results in Figures \ref{fig:w_time}c and \ref{fig:w_time}e). The local deviations are likely, as discussed above, a result of the combination of non-conserved $\mu$ and energy fluxes.

\section{Discussion}
\subsection{Interpretation of EFM energization terms}
With a better understanding of the differences between the LGCM and EFM, we are in a position to investigate the last question posed in the introduction, namely: can we associate local measurements of each EFM $W$-term with the generation of specific VDF features in a manner analogous to the betatron and Fermi terms of the LGCM? For example, one might be tempted to infer that $W_\text{diam}>0$ leads to the formation of an anisotropic thermal feature in the electron VDF. Below, we provide some arguments as to why this is not necessarily the case.

Take the dominant $W_\text{diam}=\mathbf{u}_E\cdot\nabla p_{e\perp}$ term as an example. This term describes the scalar product of the diamagnetic drift $\boldsymbol{u}_\text{diam}=-\nabla p_{e\perp} \times \boldsymbol{B}/(qnB^2)$ with $qn\boldsymbol{E}$. Unlike the curvature and magnetic gradient drifts that are present in the LGCM, the diamagnetic drift does not need to correspond to the drift of individual electrons. Rather, the pressure gradient introduces a local anisotropy in the momenta carried by gyrating electrons.
If we apply a homogeneous electric field directed along $\boldsymbol{J}_e$ over such a pressure gradient, the individual electrons will experience no net energy gain over a gyroperiod even though locally $\boldsymbol{J}_e\cdot\boldsymbol{E}>0$. What does happen, however, is that all electrons (and therefore also the pressure gradient) start $\boldsymbol{E\times B}$-drifting in the direction of the higher pressure. Thus, on a fluid level, the work done by the electric field on the diamagnetic bulk flow goes into pushing the fluid along the pressure gradient at the $\boldsymbol{E}\times\boldsymbol{B}$-velocity, as evident by the relation $qn\boldsymbol{u}_\text{diam}\cdot\boldsymbol{E}=\boldsymbol{u}_E\cdot\nabla p_{e\perp}$. 

To illustrate how the above scenario affects the local energy evolution of the plasma, we consider an Eulerian volume element that is initially in the high-pressure region. At the start the uniform plasma is $\boldsymbol{E}\times\boldsymbol{B}$-drifting across the volume element, resulting in $\nabla\cdot\boldsymbol{Q}=\partial \mathcal{E}_\text{EFM}/\partial t=\boldsymbol{J}_e\cdot\boldsymbol{E}=0$. When the pressure gradient enters the Eulerian element, however, the presence of the diamagnetic drift yields a positive $\boldsymbol{J}_e\cdot\boldsymbol{E}=W_\text{diam}$. At the same time, there is a net outward flux of high-pressure plasma that is being replaced by inflowing low-pressure plasma. In terms of the energy equation, this corresponds to $\boldsymbol{J}_e\cdot\boldsymbol{E}>0$, $\nabla\cdot\boldsymbol{Q}>0$, and $\partial\mathcal{E}_\text{EFM}/\partial t<0$. This is exactly the situation we find at the island edge around $x/d_{i0}=10$ in Figure \ref{fig:eflux}. We therefore end up with the perhaps counterintuitive conclusion that in the EFM, a positive $\boldsymbol{J}_e\cdot\boldsymbol{E}$ can lead to a decreasing energy density, due to the work being diverted into energy fluxes. We also note that the $\boldsymbol{J}_e\cdot\boldsymbol{E}>0$ region is not the source of the bulk flow, since the entire plasma is drifting, and it would do so even in the absence of the pressure gradient.

So far, we have considered an $\boldsymbol{E}$-field that is homogeneous over the electron gyro-orbits, such that no single electron gains energy during a gyroperiod in the field. If, in contrast, there are $\boldsymbol{E}$-field gradients, then the individual electrons can gain a net energy by the $\boldsymbol{E}$-field over their gyroperiod, leading to the formation of some interesting VDF feature. Importantly, however, $W_\text{diam}$ describes the local energization due to the local electric field and the local fluid current; it does not distinguish between situations where individual electrons gain energy and situations where they do not. It is therefore not possible to conclude from a local measurement of $W_\text{diam}$ whether or not individual electrons are energized, or how the VDF will change.

For completeness, a few words on what differentiates the LGCM from the EFM in this context. The reason the LGCM is able to provide insights into the evolution of the VDF (assuming $\mu$ conservation) is twofold. First, the LGCM follows the evolution of collections of guiding centers, not the evolution of the total energy content within fixed grid cells as the EFM. This means that it exclusively provides information about particle energization, and no information about energy fluxes. Second, the LGCM is already integrated over the gyroperiod, meaning that the effects of electric field gradients (or lack thereof) on the electron gyro-scales are already accounted for. There are, of course, also limitations to the electric field scales that allow for $\mu$-conservation \citep{stephens2017}, but here we assume that the LGCM is applied validly.

%\begin{figure}
%    \centering
%    \includegraphics[width=1\linewidth]{Energization_sketch_v2.pdf}
%    \caption{(a-c) The rate at which electrons gain or lose energy (red or blue respectively) depends on their position in phase space. The black lines show VDF contours for three different VDFs in an arbitrary coordinate system: (a) a Maxwellian, (b) a VDF with an asymmetry and a corresponding drift, (c) a complicated VDF with net $\boldsymbol{u}\cdot\boldsymbol{E}=0$. (d) Schematic showing that $W_\text{diam}$ need not generate energetic electrons. The electric field fills the gray shaded region. The red and blue parts of the electron orbits correspond to $q\boldsymbol{E}\cdot\boldsymbol{v}>0$ and $<0$, respectively. The $\boldsymbol{E\times B}$-drift is suppressed the sake of visual clarity. Even though $W_\text{diam}>0$ in the density gradient, the electrons experience no net energization over a full gyro orbit. (e) Same as format as (d), but $\boldsymbol{E}$ is localized to a smaller region. In this case, the electrons experience a net energy gain/loss during their interaction with the electric field.}
%    \label{fig:vdf_energization}
%\end{figure}

\subsection{Extrapolating to the physical mass ratio}

The results presented in sections \ref{sec:local} and \ref{sec:eulerian_vs_lagrangian} were obtained in the $m_i/m_e=25$ case. Similar results were also obtained for $m_i/m_e=100$. Specifically, values of $\kappa<1$ and large errors in the guiding center model at the x-points and near the current sheet center were also found in the $m_i/m_e=100$, $B_g=0$ case. How these results change for more realistic mass ratios remains to be investigated. It is possible that an increased mass ratio could affect the relative importance of the $W$-terms. However, as mentioned in the above paragraphs, the EFM energization terms do not necessarily reflect the evolution of the VDF, and any application of a decomposition of $\boldsymbol{J}_e\cdot\boldsymbol{E}$ into the EFM $W$-terms must be designed accordingly.  
Moreover, accurately computing the EFM terms using in-situ data is a challenging task since, in addition to the $p_\perp$ and $p_\parallel$ required by the LGCM, the EFM also requires accurate measurements of pressure gradients in $W_\text{diam}\propto\nabla p_{e\perp}$ and $W_\text{agyro}\propto\nabla\cdot\boldsymbol{\Uppi}^a_e$.

One important result from our analysis is that the LGCM can produce significant errors due to the erroneous assumption of $\mu$-conservation when $\kappa<1$. In our simulations, we found regions of low $\kappa$ at the x-points, near the island edges, and inside the islands for $B_g=0$ (Figure \ref{fig:guiding_error}d). Increasing the guide field to $0.2$ increased $\kappa$, particularly inside the islands, but low-$\kappa$ regions still remained. Since the reconnecting current sheet is on electron-scales, increasing the mass-ratio should not affect $\kappa$ near the current sheet center or in the electron diffusion region. Indeed, from spacecraft observations it is well known that electrons demagnetize in these regions \citep[e.g.][]{burch2016,torbert2018,cozzani2019}. However, since the magnetic islands grow into ion-scale structures, the associated curvature radius should increase for increasing mass-ratios. We might therefore expect that the LGCM problem of demagnetized electrons at magnetic islands should be less of a concern for physical mass-ratios. However, the islands still grow from electron scales, so this argument would only be valid in the later stages of reconnection. How well the electron magnetic moment is conserved may also depend on several other parameters like the plasma beta and the current sheet type (Harris versus force-free). These parameters can therefore affect the validity of the LGCM. In any case, we know from spacecraft data that electron scale structures with strong gradients can be present in physical reconnection outflows in geospace \citep[e.g.][]{zhou2019,leonenko2021}. It is therefore important to be aware of the fact that the LGCM assumes $\mu$ conservation, and to test that assumption before applying the model.

\section{Conclusions}
In the present paper, we build on the recent work by \citet{tenbarge2024} to study the Eulerian fluid model (EFM; Eq.\ (\ref{eq:fluid})) and the Lagrangian guiding center model (LGCM; Eq.\ (\ref{eq:guiding})) of electron energization during multiple x-point reconnection using particle-in-cell simulations.
We set out to answer the series of questions summarized below.
\begin{enumerate}
    \item Which EFM energization terms are most important during multi-x-point reconnection, and are they affected by the guide field ($B_g$) or the ion-to-electron mass ratio?
    \item What is the reason for the different predictions by the LGCM and EFM, and what does that tell us about the electrons?
    \item The energization terms in the LGCM are often interpreted as leading to specific changes in the electron velocity distribution function (VDF). Can we make similar interpretations for the EFM energization terms?
\end{enumerate}

(i) We find that the energization related to the diamagnetic drift ($W_\text{diam}$) is generally, on large scales, the dominant contributor to $\boldsymbol{J}_e\cdot\boldsymbol{E}$ in the EFM (see Figure \ref{fig:w_time}). For large guide fields ($B_g=1$), the contribution of parallel electric fields, $W_\parallel$, becomes the most important term (albeit still comparable to $W_\text{diam}$). The increased importance of $W_\parallel$ for larger $B_g$ has previously been observed for the LGCM \citep{dahlin2016}. The only clear mass-ratio dependence is found for the Eulerian curvature term ($W_\text{curv1}$) in the $B_g=0$ case, where it becomes less important as $m_i/m_e$ is increased from 25 to 100. Locally, other energization terms can contribute significantly to electron energization (see Figure \ref{fig:local_energization}).

(ii) We show that the assumption of magnetic moment conservation made by the LGCM is invalid in the low $B_g\leq0.2$ case near the x-points and at the center of the current sheets where the magnetic field is strongly curved. This causes significant errors in the LGCM energization estimate. For $B_g=1$, the electrons remain magnetized, and we find a good agreement between the simulation domain integrated LGCM and $\boldsymbol{J}_e\cdot\boldsymbol{E}$. Local deviations from $\boldsymbol{J}_e\cdot\boldsymbol{E}$ are produced by the LGCM in all simulations, even when $\mu$ is conserved. This is due to the fact that the LGCM and the EFM describe two fundamentally different quantities. The LGCM describes the evolution of the total energy of the guiding centers that are currently within a given volume element.
In contrast, the EFM describes the evolution of the total electron energy content within the volume element.
The latter includes the effects of energy fluxes through the volume element, whereas the former does not. This has the consequence that the two models can both be correct, while simultaneously giving seemingly contradictory local energization rates if there is a net energy flux into or out of the volume element under consideration. By examining the different terms of the electron energy equation in Figure \ref{fig:eflux}, we find that the energy flux terms are important in balancing $\boldsymbol{J}_e\cdot\boldsymbol{E}$ on a local level, explaining the difference between the two models when $\mu$ is conserved.

(iii) Finally, we argue that, unlike the LGCM, we cannot associate local measurements of the different EFM energization terms with the formation of structures in the electron VDF. In essence, this is due to the fact that the EFM includes the local effect of energy fluxes and fluid drifts which need not correspond to the drift of individual particles.

In summary, the EFM describes a complete decomposition of $\boldsymbol{J}_e$ into fluid drifts, and it therefore accurately reproduces $\boldsymbol{J}_e\cdot\boldsymbol{E}$. Local measurements of each constituent energization term, however, need not describe locations where the plasma is either heated or accelerated. In contrast, the LGCM accurately describes the energization of particles if it is used validly, i.e.\ when $\mu$ is conserved, but it does not describe the local evolution of the bulk fluid energy. Local deviations between the energization measures of the two models, even when $\mu$ is conserved, can be attributed to the contrasting Eulerian and Lagrangian perspectives, and does not mean that one of the models is wrong. A thorough understanding of what one wishes to describe is therefore necessary before one decides to use either of the energization models.

\section*{Acknowledgments}
We are grateful to J. M. TenBarge for helpful correspondence at an early stage of this work. 
This project received funding from the Knut and Alice Wallenberg Foundation (Grant No. KAW 2022.0087) and the Swedish Research Council (Dnr.~2021-03943). KS acknowledges financial support from the Adlerbertska Research Foundation. The computations and data handling were enabled by resources provided by the National Academic Infrastructure for Supercomputing in Sweden (NAISS), partially funded by the Swedish Research Council through grant agreement no. 2022-06725.

\appendix

\section{Generalized adiabaticity estimate}
\label{gradBmeasure}

\begin{figure}
    \centering
    \includegraphics[width=1\linewidth]{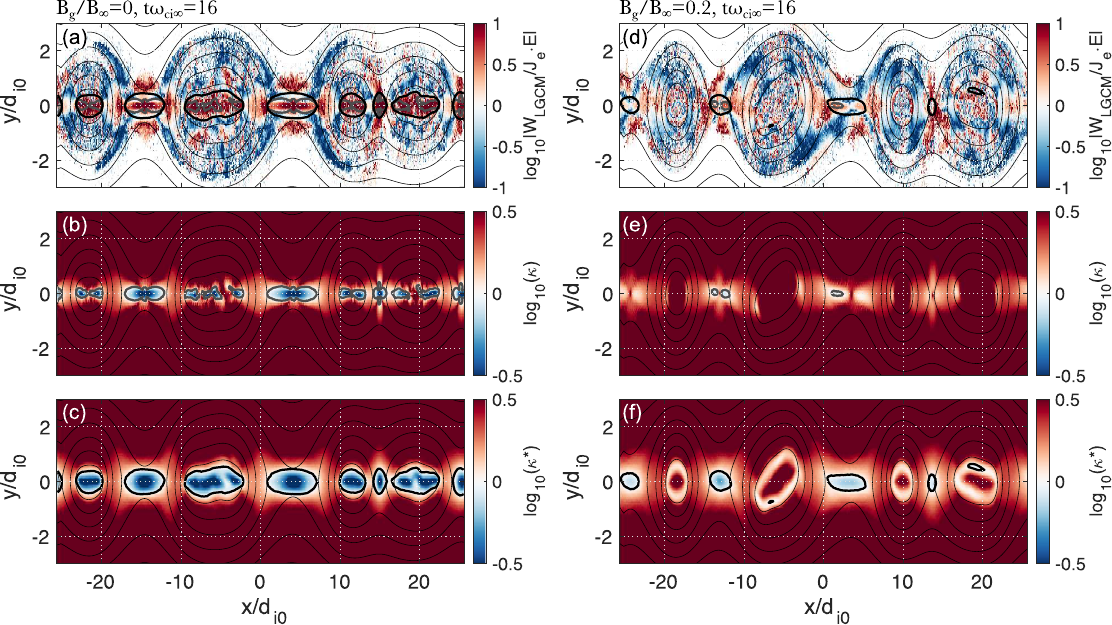}
    \caption{Difference between $\kappa$ and $\kappa^*$ for $B_g/B_\infty=0$ (left column) and $B_g/B_\infty=0.2$ (right column). (a) Ratio between the LGCM sum and $\boldsymbol{J}_e\cdot\boldsymbol{E}$. The thick black and grey contours correspond to $\kappa^*=1$ and $\kappa=1$, respectively. (b) $\kappa$ calculated using $r_\text{curv}$. The thick grey contours correspond to $\kappa=1. $(c) $\kappa^*$ calculated using $L_{\nabla\boldsymbol{B}}$. The thick black contours correspond to $\kappa^*=1$. (d-f) Same format as panels (a-c). All colormaps are saturated.}
    \label{fig:kappa_star}
\end{figure}

The standard adiabaticity parameter $\kappa=\sqrt{r_\text{curv}/r_{ge}}$ uses the magnetic field curvature radius \citep{buchner1989}, $r_\text{curv}=1/|\boldsymbol{b}\cdot\nabla\boldsymbol{b}|$, to quantify the adiabaticity, and does not account for the scales of magnetic field gradients. A generalized length scale that incorporates all magnetic field variations can be constructed as $L_{\nabla\boldsymbol{B}}^{-1}=\Vert \nabla\boldsymbol{B}/B\Vert$, where $\Vert\cdot\Vert$ denotes the spectral norm. This length scale captures both curvature and gradient effects. We can then define an adiabaticity measure analogous to $\kappa$ as $\kappa^*=\sqrt{L_{\nabla\boldsymbol{B}}/r_{ge}}$. Figure \ref{fig:kappa_star} shows that the regions where $\kappa$ predicts non-adiabatic electrons ($\kappa<1$) are confined within regions of $\kappa^*<1$, as expected. Because the magnetic field is strongly curved, the two quantities are similar, and the $\kappa^*<1$ regions only extend slightly outside the $\kappa<1$ regions in the $B_g=0$ case (Figures \ref{fig:kappa_star}b, \ref{fig:kappa_star}c). We note that the $\nabla \cdot \mathbf{B}=0$ constraint makes field strength variations correlated with curvature, so this result is not unexpected. In the $B_g=0.2$ case, the generalized $\kappa^*$ picks up potentially non-adiabatic regions at island boundaries ($\kappa^*\approx1)$, which are not captured by $\kappa$. These are regions with strong magnetic field gradients. 

In summary, other length and time scales can also affect the adiabaticity of electrons \citep{stephens2017}. However, $\kappa$ is sufficient to show that electrons can be non-adiabatic in both the low and intermediate guide field cases, and that $\kappa<1$ correlates strongly with large deviations between the LGCM sum and $\boldsymbol{J}_e\cdot\boldsymbol{E}$.

\section{Local energization and demagnetization for $B_g=1$}
\label{app:bg1}

\begin{figure}
    \centering
    \includegraphics[width=1\linewidth]{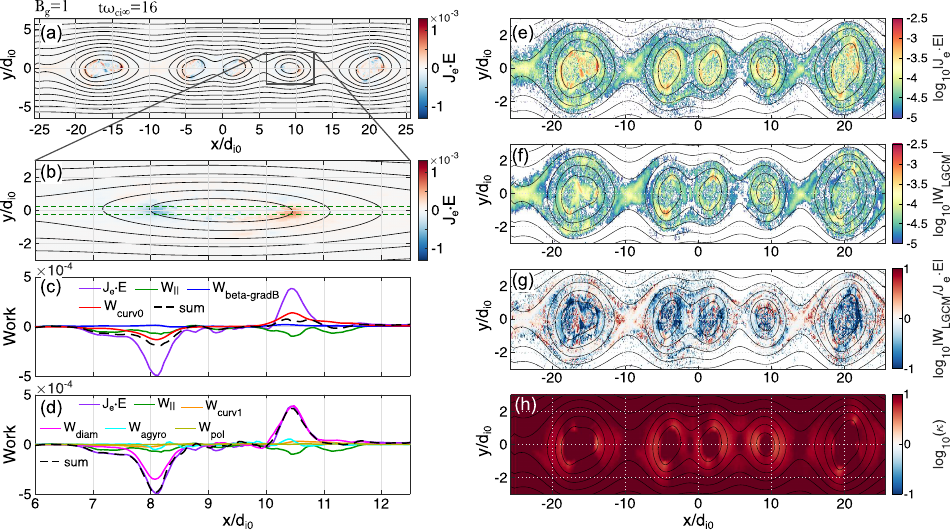}
    \caption{(a-d) Spatial dependence of $\boldsymbol{J}_e\cdot\boldsymbol{E}$ and the $W$ terms inside magnetic islands for $B_g=1$. Same format as Figure \ref{fig:local_energization}. (e-h) Deviation between the LGCM sum and $\boldsymbol{J}_e\cdot\boldsymbol{E}$ for $B_g=1$. Same format as Figure \ref{fig:guiding_error}. Note that we have increased the range of the colormap in (h) compared to Figure \ref{fig:guiding_error}, as $\log_{10}(\kappa)>0.5$ everywhere in this case.}
    \label{fig:bg1}
\end{figure}

In this appendix, we present the data corresponding to Figures \ref{fig:local_energization} and \ref{fig:guiding_error} for the $B_g=1$ case in Figure \ref{fig:bg1}. 

Figure \ref{fig:bg1}c shows that the sum of the LGCM energization terms deviates significantly from $\boldsymbol{J}_e\cdot\boldsymbol{E}$ in the $B_g=1$ case. The EFM (Figure \ref{fig:bg1}d), on the other hand, provides an accurate estimate of $\boldsymbol{J}_e\cdot\boldsymbol{E}$. 
We note that the reversed polarity of $\boldsymbol{J}_e\cdot\boldsymbol{E}$ compared to the $B_g=0$ and $B_g=0.2$ cases (see Figure \ref{fig:local_energization}) is due to the fact that the island in the $B_g=1$ case is convecting in the opposite ($-x$) direction. 

As shown in Figure \ref{fig:bg1}g, the LGCM tends to provide an energization estimate that is much less than $|\boldsymbol{J}_e\cdot\boldsymbol{E}|$, particularly inside the magnetic islands. These deviations cannot be explained by electron demagnetization as quantified by $\kappa$, since $\kappa\gg1$ throughout the whole simulation domain (Figure \ref{fig:bg1}h).

\bibliographystyle{apalike}
\bibliography{biblio.bib}

\end{document}